\documentclass[12pt]{article}
\begin{document}
\title{Noncommutative Gauge Fields and Mass Generation}
\author{B.G. Sidharth\\
International Institute for Applicable Mathematics \& Information Sciences\\
Hyderabad (India) \& Udine (Italy)\\
B.M. Birla Science Centre, Adarsh Nagar, Hyderabad - 500 063 (India)}
\date{}
\maketitle
\begin{abstract}
It is well known that a typical Yang-Mills Gauge Field is mediated by massless Bosons. It is only through a symmetry breaking mechanism, as in the Salam-Weinberg model that the quanta of such an interaction field acquire a mass in the usual theory. Here we demonstrate that without taking recourse to the usual symmetry breaking mechanism, it is possible to have massive Gauge Fields, given a noncommutative geometrical underpinning for spacetime.
\end{abstract}
\section{Introduction}
For nearly seventy five years there have been fruitless attempts at unifying electromagnetism and gravitation, starting with the very early attempt of Hermann Weyl and his gauge invariant geometry \cite{weyl}. These attempts later evolved into Quantum Gravity schemes on the one hand, while independently Quantum Super String theory also has shown promise in this direction \cite{garay,amati,ven,bgsgrav,cu}. Curiously enough Weyl's original ideas in Classical Physics were developed decades later in a Quantum Mechanical context to evolve into the Yang-Mills gauge theory and subsequent developments \cite{lee,taylor,creutz}.\\
It is of course well known that the gauge theory approach has proved most fruitful in the case of the electroweak interactions and has also made inroads into the domain of strong interactions. But it has fallen short of gravitation. A notable feature of the modern non-Abelian Gauge theories is the mass generation, for which recourse has to be taken through a symmetry breaking mechanism.\\
Returning to the case of gravitation, a notable feature of the recent schemes is the existence of a minimum spacetime cut off. Indeed as t' Hooft has remarked\cite{hooft}: ``It is some what puzzling to the present author why the lattice structure of space and time has escaped attention from other investigators up till now.'' One of the consequences of such a minimum cut off is that there is an underlying noncommutative geometric structure of spacetime (Cf.refs. \cite{cu,bgsmup} and other references therein). This is also symptomatic of the fact that the underlying spacetime is no longer a differenciable manifold. We will now argue that the noncommutative geometry is a mechanism which generates mass in a non-Abelian Gauge Field theory.
\section{Noncommutative Non Abelian Gauge Fields}
Let us now consider the Gauge Field itself. As is well known, this could be obtained as a generalization of the above phase function $\lambda$ to include fields with internal degrees of freedom. For example $\lambda$ could be replaced by $A_\mu$ given by 
\begin{equation}
A_\mu = \sum_{\imath} A^\imath_\mu (x)L_\imath ,\label{e1}
\end{equation}
The Gauge Field itself would be obtained by using Stoke's Theorem and (\ref{e1}). This is a very well known procedure: considering a circuit, which for simplicity we can take to be a parallellogram of side $dx$ and $dy$ in two dimensions, we can easily deduce the equation for the field, viz.,
\begin{equation}
F_{\mu \nu} = \partial_\mu A_\nu - \partial_\nu A_\mu - \imath q [A_\mu , A_\nu ],\label{e2}
\end{equation}
$q$ being the Gauge Field coupling constant.\\
In (\ref{e2}), the second term on the right side is typical of a non Abelian Gauge Field. In the case of the $U(1)$ electromagnetic field, this latter term vanishes.\\
Infact as is well known, in a typical Lagrangian like 
\begin{equation}
\mathit{L} = \imath \bar \psi \gamma^\mu D_\mu \psi - \frac{1}{4} F^{\mu \nu} F_{\mu \nu} - m \bar \psi \psi\label{e3}
\end{equation}
$D$ denoting the Gauge covariant derivative, there is no mass term for the field Bosons. Such a mass term in (\ref{e3}) must have the form $m^2 A^\mu A_\mu$ which unfortunately is not Gauge invariant.\\
This was the shortcoming of the original Yang-Mills Gauge Theory: The Gauge Bosons would be massless and hence the need for a symmetry breaking, mass generating mechanism.\\
The well known remedy for the above situation has been to consider, in analogy with superconductivity theory, an extra phase of a self coherent system (Cf.ref.\cite{moriato} for a simple and elegant treatment). Thus instead of the Gauge Field $A_\mu$, we consider a new phase adjusted Gauge Field after the symmetry is broken
\begin{equation}
W_\mu = A_\mu - \frac{1}{q} \partial_\mu \phi\label{e4}
\end{equation}
The field $W_\mu$ now generates the mass in a self consistent manner via a Higgs mechanism. Infact the kinetic energy term
\begin{equation}
\frac{1}{2} |D_\mu \phi |^2\quad ,\label{e5}
\end{equation}
where $D_\mu$ in (\ref{e5})denotes the Gauge covariant derivative, now becomes
\begin{equation}
|D_\mu \phi_0 |^2 = q^2|W_\mu |^2 |\phi_0 |^2 \, ,\label{e6}
\end{equation}
Equation (\ref{e6}) gives the mass in terms of the ground state $\phi_0$.\\
Let us now consider in the Gauge Field transformation, an additional phase term, $f(x)$, this being a scalar. In the usual theory such a term can always be gauged away in the $U(1)$ electromagnetic group. However we now consider the new situation of a noncommutative geometry referred to above, 
\begin{equation}
\left[dx^\mu , dx^\nu \right] = \Theta^{\mu \nu} \beta , \beta \sim 0 (l^2)\label{e7}
\end{equation}
where $l$ denotes the minimum spacetime cut off. (Cf. also ref.\cite{nc,annales}). Then the $f$ phase factor gives a contribution to the second order in coordinate differentials,
$$\frac{1}{2} \left[\partial_\mu B_\nu - \partial_\nu B_\mu \right] \left[dx^\mu , dx^\nu \right]$$
\begin{equation}
+ \frac{1}{2} \left[\partial_\mu B_\nu + \partial_\nu B_\mu \right] \left[dx^\mu dx^\nu + dx^\nu dx^\mu \right]\label{e8}
\end{equation}
where $B_\mu \equiv \partial_\mu f$.\\
As can be seen from (\ref{e8}) and (\ref{e7}), the new contribution is in the term which contains the commutator of the coordinate differentials, and not in the symmetric second term. Effectively, remembering that $B_\mu$ arises from the scalar phase factor, and not from the non-Abelian Gauge Field, in equation (\ref{e2}) $A_\mu$ is replaced by 
\begin{equation}
A_\mu \to A_\mu + B_\mu = A_\mu + \partial_\mu f\label{e9}
\end{equation}
Comparing (\ref{e9}) with (\ref{e4}) we can immediately see that the effect of noncommutativity is precisely that of providing a Gauge invariant mass term to the Gauge Field.\\
On the other hand if we neglect in (\ref{e7}) terms $\sim l^2$, then there is no extra contribution coming from (\ref{e8}) or (\ref{e9}), so that we are in the usual non-Abelian Gauge Field theory, requiring a broken symmetry to obtain an equation like (\ref{e9}). This is not surprising because if we neglect term $\sim l^2$ in (\ref{e7}) then we are back with the usual commutative theory and the usual Quantum Mechanics.\\
Let us now consider the symmetric term in (\ref{e8}). This is equivalent to retaining terms $\sim l^2$, that is squares of the coordinate differentials. Thus the phase transformation with $f$ gives a term like
\begin{equation}
\left\{ \partial_\mu f \right\} dx^\mu + \left(\partial_\mu \partial_\nu + \partial_\nu \partial_\mu \right) f \cdot dx^\mu dx^\nu\label{e10}
\end{equation}   
We must remember that neither the derivatives nor the products of coordinate differentials now commute.\\
As in the usual theory the coefficient of $dx^\mu$ in the first term of (\ref{e10}) represents now, not the gauge term but the electromagnetic potential itself: Infact, in this noncommutative geometry, it can be shown that this electromagnetic potential reduces to the potential in Weyl's original gauge theory.\\
Without the noncommutativity, the potential $\partial_\mu f$ would lead to a vanishing electromagnetic field. However Dirac pointed out in his famous monopole paper in 1930 that a non integrable phase $f (x,y,z)$ leads as above directly to the electromagnetic potential, and moreover this was an alternative formulation of the original Weyl theory.\\
Returning to (\ref{e10}) we identify the next coefficient with the metric tensor giving the gravitational field:
\begin{equation}
ds^2 = g_{\mu \nu} dx^\mu dx^\nu = \left(\partial_\mu \partial_\nu + \partial_\nu \partial_\mu \right) f dx^\mu dx^\nu\label{e11}
\end{equation}
Infact one can easily verify that $ds^2$ of (\ref{e11}) is an invariant. We now specialize to the case of the linear theory in which squares and higher powers of  $h^{\alpha \beta}$ can be neglected. In this case it can easily be shown that
\begin{equation}
2 \Gamma^\beta_{\mu \nu} = h_{\beta  \mu ,\nu} + h_{\nu \beta ,\mu} - h_{\mu \nu ,\beta}\label{e12}
\end{equation}
where in (\ref{e12}), the $\Gamma$s denote Christofell symbols. From (\ref{e12}) by a contraction we have
\begin{equation}
2\Gamma^\mu_{\mu \nu} = h_{\mu \nu ,\mu} = h_{\mu \mu , \nu}\label{e13}
\end{equation}
If we use the well known gauge condition 
$$\partial_\mu \left(h^{\mu \nu} - \frac{1}{2} \eta^{\mu \nu} h_{\mu \nu}\right) = 0, \, \mbox{where}\, h = h^\mu_\mu$$
then we get
\begin{equation}
\partial_\mu h_{\mu \nu} = \partial_\nu h^\mu_\mu = \partial_\nu h\label{e14}
\end{equation}
(\ref{e14}) shows that we can take the $f$ in (\ref{e10}) as $f = h$, both for the electromagnetic potential $A_\mu$ and the metric tensor $h_{\mu \nu}$. (\ref{e13}) further shows that the $A_\mu$ so defined becomes identical to Weyl's gauge invariant potential.\\
However it is worth reiterating that in the present formulation, we have a noncommutative geometry, that is the derivatives do not commute and moreover we are working to the order where $l^2$ cannot be neglected. Given this condition both the electromagnetic potential and the gravitational potential are seen to follow from the gauge like theory. By retaining coordinate differential squares, we are even able to accommodate apart from the usual spin 1 gauge particles, also the spin 2 graviton which otherwise cannot be accommodated in the usual gauge theory. If however $O(l^2) = 0$, then we are back with commutative spacetime, that is a usual point spacetime and the usual gauge theory describing spin 1 particles.
\section{Noncommutativity and the Modified Uncertainty Principle}
It is well known that the noncommutativity in (\ref{e7}) leads to a modification of the usual Uncertainty Principle, which now becomes
\begin{equation}
\Delta x \sim \frac{\hbar}{\Delta p} + \alpha' \frac{\Delta p}{\hbar}\label{e15}  
\end{equation}
where $\alpha' = l^2$. This is an expression of a duality relation
$$R \to \alpha' /R,$$
and is symptomatic of the fact that we cannot go down to arbitrarily small spacetime intervals, but that the macro universe is connected with the micro universe. As Witten put it \cite{witten}, ``when one accelerates past the string scale - instead of probing short distances, one just watches the propagation of large strings.''\\
To see in greater detail, this connection between the small and large scales, we now use the fact that $\sqrt{N}$ is the fluctuation in the number of particles, $N \sim 10^{80}$, in the universe. So
$$\Delta p = \sqrt{N}mpc$$
is the fluctuation in the momentum, $m$ being the mass of a typical elementary particle. Using this in the second or extra Uncertainty term on the right side of (\ref{e15}) we get,
\begin{equation}
R = \sqrt{N} l\label{e16}
\end{equation}
This is the well known so called Eddington relation giving the radius of the universe in terms of a typical Compton wavelength $l$. One could now invert the picture and start, not with the modified Uncertainty relation, but rather with the universe at large. It is well known that in a gas containing $N$ particles and of extension $R$, the Uncertainty in the position of the particle, $l$ is given precisely by (\ref{e16}), through the Brownian Motion of the constituents. Infact this dependence of Quantum Mechanics on cosmic fluctuations has been discussed earlier \cite{bgscos}. All this goes to show that the many supposedly miraculous and inexplicable large number coincidences which we encounter in cosmology are infact consequences of this deeper principle in action. Such a cosmology was worked out and successfully predicted as accelerating ever expanding universe with dark energy content \cite{ijmpa}.


\begin{thebibliography}{99}
\bibitem {weyl}H. Weyl, {\it Space-Time Matter} (Denver Publications
Inc.,New York, 1962), 282ff.
\bibitem {garay} L.J. Garay, {\it Int.J.Mod.Phys. A}, Vol.10, No.2, 1995, p.145-165.
\bibitem {amati} D. Amati in {\it Sakharov Memorial Lectures}, Eds. L.V. Kaddysh
and N.Y. Feinberg, (Nova Science, New York, 1992), pp.455ff.
\bibitem {ven} G. Veneziano in "The Geometric Universe", Ed. by
S.A. Huggett, et al.,  Oxford University Press, Oxford, 1998, p.235ff.
\bibitem {bgsgrav} B.G. Sidharth, Gravitation and Cosmology, 4 (2) (14), 1998, p.158ff.
\bibitem {cu} B.G. Sidharth, "Chaotic Universe: From the Planck to the Hubble Scale", Nova Science Publishers, Inc., New York, 2001.
\bibitem {lee} B. Lee and W. Abers in ``Gauge Theory and Neutrino Physics'', Ed., M. Jacob, North Holland, Amsterdam, 1978.
\bibitem {taylor} J.C. Taylor, ``Gauge Theories of Weak Interactions'', Cambridge University Press, London, 1976.
\bibitem {creutz} M. Creutz, ``Quarks, Glucons and Lattices'', Cambridge University Press, London, 1983.
\bibitem {hooft} G. 't Hooft, ArXiv:gr-qc/9601014.
\bibitem {bgsmup} B.G. Sidharth, Chaos, Solitons and Fractals, 15, 2003, 593-595.
\bibitem {moriato} K. Moriyasu, ``An Elementary Primer for Gauge Theory'', World Scientific, Singapore, 1983.
\bibitem {nc} B.G. Sidharth, Il Nuovo Cimento, 117B (6), 2002, 703ff.
\bibitem {annales}  B.G Sidharth, Annales de la Fondation Louis de Broglie, 27 (2), 2002, pp.333ff.
\bibitem {witten} W. Witten, Physics Today, April 1996, pp.24-30.
\bibitem {bgscos} B.G. Sidharth, Chaos, Solitons and Fractals, 15 (1), 2003, pp.25-28.
\bibitem {ijmpa} B.G. Sidharth, Int.J.Mod.Phys.A, 13 (15), 1998, p.2599ff.
\end{thebibliography}
\end{document}